Author for correspondence:
G. A. D. Briggs
e-mail: andrew.briggs@materials.ox.ac.uk

# The Oxford Questions on the foundations of quantum physics

G. A. D. Briggs[1], J. N. Butterfield[2] and A. Zeilinger[3]

[1]Department of Materials, University of Oxford, Oxford OX1 3PH, UK
[2]Trinity College, University of Cambridge, Cambridge CB2 1TQ, UK
[3]Faculty of Physics, University of Vienna, Vienna, Austria

The twentieth century saw two fundamental revolutions in physics—relativity and quantum. Daily use of these theories can numb the sense of wonder at their immense empirical success. Does their instrumental effectiveness stand on the rock of secure concepts or the sand of unresolved fundamentals? Does measuring a quantum system probe, or even create, reality or merely change belief? Must relativity and quantum theory just coexist or might we find a new theory which unifies the two? To bring such questions into sharper focus, we convened a conference on Quantum Physics and the Nature of Reality. Some issues remain as controversial as ever, but some are being nudged by theory's secret weapon of experiment.

## 1. The achievements of twentieth century physics

Much of the history of twentieth century physics is the story of the consolidation of the relativity and quantum revolutions, with their basic postulates being applied ever more widely. It is possible to forget how contingent, indeed surprising, it is that the basic postulates of relativity and quantum theory have proved to be so successful in domains of application far beyond their original ones. Why should the new chronogeometry, introduced by Einstein's special relativity in 1905 [1] for electromagnetism, be extendible to mechanics, thermodynamics and other fields of physics? And why should the quantum theory devised for systems of atomic dimensions ($10^{-10}$ m) be good for scales both much smaller (cf. high-energy experiments $10^{-17}$ to $10^{-20}$ m) and vastly larger (cf. superconductivity and superfluidity, or even a neutron interferometer, involving scales of a fraction of a metre or more)? Is there an upper limit to the scale on which quantum theory should be expected to work? There is a sense in which all properties of matter are quantum mechanical. Topics as diverse as phase changes of alloys and conduction in semiconductors have all yielded to quantum theory. New quantum mechanical models are being developed for a growing range of superconductors, magnets, multiferroics and topological insulators.

The point applies equally well when we look beyond terrestrial physics. General relativity makes a wonderful story: the theory was created principally by one person, motivated by conceptual, in part genuinely philosophical, considerations—yet, it has proved experimentally accurate in all kinds of astronomical situations. They range from weak gravitational fields such as occur in the solar system, where it famously explains the minuscule precession of the perihelion of Mercury (43″ of arc per century) that was unaccounted for by Newtonian theory, to fields 10 000 times stronger in a distant binary pulsar, which in the last 30 years has given us compelling evidence for a phenomenon (gravitational radiation) that was predicted by general relativity and long searched for.

But general relativity is not the only success story in modern physics' description of non-terrestrial phenomena. Quantum theory has also been extraordinarily successful in application to astronomy: the obvious example is the use of nuclear physics to develop a very accurate and detailed theory of stellar structure and evolution.

Indeed, there is a more general point here, going beyond the successes of relativity and quantum theory. We tend to get used to the various unities in Nature that science reveals—and thereby to forget how contingent and surprising they are. Of course, this is not just a tendency of our own era. For example, nineteenth century physics confirmed Newton's law of gravitation to apply outside the Solar System and discovered terrestrial elements to exist in the stars (by spectroscopy): discoveries that were initially surprising, but soon taken for granted, incorporated into the educated person's 'common sense'. Similarly, nowadays: the many and varied successes of physics in the last few decades, in modelling very accurately phenomena that are vastly distant in space and time, and/or very different from our usual 'laboratory scales' (in their characteristic values of such quantities as energy, temperature or pressure, etc.), reveal an amazing unity in Nature. For a modern, specific (and literally spectacular) example, consider the precision and detail of our models of supernovae—as confirmed by the wonderful capacity of modern telescope technology to see and analyse individual supernovae, even in other galaxies.

In some cases, a theoretical prediction came first, stimulating experimental confirmation; sometimes, it has been the other way round. It was nearly half a century after the experimental discovery of superconductivity that a satisfactory quantum theory was developed, but the prediction of the Josephson junction preceded its experimental implementation. There had long been a quest for higher temperature superconductors, but their remarkable discovery came as rather astonishing.

And yet: complacency, let alone triumphalism, is not in order! Not only is physics full of unfinished business: that is always true in human enquiry. We think most physicists would also agree that there are clouds on the horizon that may prove as great a threat to the extrapolated success of twentieth century physics, in particular quantum physics, as were the anomalies confronting classical physics at the end of the nineteenth century. But physicists might well disagree about what these clouds are in broad terms as well as in their details.

In 2010, believing that the time was ripe for better defining these clouds, we organized a conference entitled Quantum Physics and the Nature of Reality. In §2, we describe it, and how it led to the formulation of a list of main open questions about the foundations of quantum physics. In §3, we discuss progress on the questions; and in §4, we return to more general discussion of the present situation in physics.

## 2. The Oxford conference on Quantum Physics and the Nature of Reality

The conference was a celebration of the 80th birthday of Reverend Dr John Polkinghorne, in recognition of his wide-ranging enquiries into the deeper significance of quantum physics,

> **Box 1.** The Oxford Questions.
>
> (1) Time, irreversibility, entropy and information
>    (a) Is irreversibility fundamental for describing the classical world?
>    (b) How is irreversibility involved in quantum measurement?
>    (c) What can we learn about quantum physics by using the notion of information?
>
> (2) The quantum–classical relationships
>    (a) Does the classical world emerge from the quantum, and if so which concepts are needed to describe this emergence?
>    (b) How should we understand the transition from observation to informed action?
>    (c) How can a single-world realistic interpretation of quantum theory be compatible with non-locality and special relativity?
>
> (3) Experiments to probe the foundations of quantum physics
>    (a) What experiments can probe macroscopic superpositions, including tests of Leggett–Garg inequalities?
>    (b) What experiments are useful for large complex systems, including technological and biological?
>    (c) How can the progressive collapse of the wave function be experimentally monitored?
>
> (4) Quantum physics in the landscape of theories
>    (a) What insights are to be gained from category-theoretic, informational, geometric and operational approaches to formulating quantum theory?
>    (b) What are productive heuristics for revisions of quantum theory?
>    (c) How does quantum physics cohere with space–time and with mass–energy?
>
> (5) Interaction with questions in philosophy
>    (a) How do different aspects of the notion of reality influence our assessment of the different interpretations of quantum theory?
>    (b) How do different concepts of probability contribute to interpreting quantum theory?

building on his distinguished career in mathematical quantum theory. The participants included, in roughly equal proportions: experimentalists, theoreticians and philosophers of physics. The conference was carefully planned to produce a set of questions, to be known as the Oxford Questions, which would gather the collected wisdom of all three disciplines in a form which would be both far-reaching and tractable.

From all three viewpoints—experimental, theoretical and philosophical—the foundations of quantum physics is a thriving, lively and even controversial field of research. Across the world, many active groups of researchers are engaged in probing the nature of quantum theory— and the nature of reality, as described by quantum theory. This work draws—equally, and to a considerable extent, synergistically—on the expertise of the three communities. Accordingly, in organizing the conference, we were keen to invite a broad range of speakers from the various facets of the subject. But we also wanted to avoid re-hashing various aspects of the status quo in debates about the foundations of quantum physics: we instead aspired to identify (albeit contentiously!) a set of central open problems about the nature of quantum reality, to stimulate and guide future research and scholarship. To this end, we

(i) invited some participants (again: including experimentalists, theoreticians and philosophers) to write a short white paper which was circulated in advance;

(ii) asked speakers to zoom out from the details of their latest research and give very short talks; and
(iii) asked speakers, in discussion periods and coffee-breaks, to meet with us to help formulate the open problems—which, by the end of the conference, we collectively settled on as the *Oxford Questions* (see box 1).

## 3. The questions in play

For all three communities in quantum foundations—experimentalists, theoreticians and philosophers—the last 20 or so years have been especially rich, thanks to the rise of quantum information science (taken as including quantum computation and quantum cryptography). This has stimulated, and been stimulated by, extraordinary advances in experimental progress in a range of implementations across many fields of physics: in optics, trapped ions and lattices of atoms, in three different ways in superconductors (viz. phase, flux and Cooper-pair box), and in nuclear and electron spins in molecules, semiconductors and diamond.

Another prominent theme for all three communities, throughout the last 20 years, has been decoherence: that is, the extremely fast and ubiquitous process by which information about the quantum system propagates into the environment, so that we lose the ability to show quantum interference. The interaction between the system and the environment is normally such as to leave the system in an improper mixture whose density matrix is nearly diagonal in a quantity; for example, the position of the centre of mass that we intuitively wish to be definite in value.

Here follows a brief general description of activities over recent decades for each of the three communities.

### (a) Theory

Ever since the inception of quantum theory, theoreticians have successively deepened our understanding of its conceptual and mathematical structure. Confining ourselves to the growth in foundational studies since the 1960s, any list of highlights must surely include

(i) the discovery of the various Bell inequalities, the Bell–Kochen–Specker theorem, the Leggett–Garg inequality;
(ii) various deeper analyses, e.g. of quantization, of uncertainty relations, of the convex set structure of quantum state spaces, of positive operator valued measures (also known as: operational quantum physics);
(iii) the development of alternatives to quantum theory, such as the dynamical reduction models of such authors as Ghirardi, Pearle, Penrose and Adler, whose differences from quantum theory may in the foreseeable future be subjected to experimental test; and
(iv) with the rise of quantum information theory: analyses using ideas from such diverse fields as information theory, complexity theory (applied to communication, computation and cryptography) and category theory.

### (b) Experiment

Experimentalists have in the last 30 or so years performed in their laboratories many of the Gedanken experiments on single quantum systems which were first proposed by the founding fathers of quantum theory. These experiments, together with many other foundational experiments, for example, on multiphoton entanglement, or those that test Bell, Bell–Kochen–Specker or Leggett-type inequalities, have served to make the interpretative issues about the theory more vivid [2]. All the more so, with the rise of quantum information theory.

(i) The Bell inequality has been experimentally violated in a number of implementations [3], thus experimentally demonstrating entanglement and thus showing that quantum mechanics is not incomplete in the sense envisaged by Einstein *et al.* [4] in 1935.

Various delayed-choice experiments have been carried out, successively ruling out any reconciliation of Bohr's complementarity with Einstein's local conception of physical reality [5], and the experiments have been extended with increasing degrees of sophistication to multi-particle situations [6].

(ii) Quantum interference has been demonstrated in diffraction experiments with molecules of increasing size, most recently with molecules containing up to 430 atoms [7]. This is not yet large enough to test continuous spontaneous localization models, but it does show that no deviation from quantum predictions is found up to this scale.

(iii) The Leggett–Garg inequality has been tested in various photonic, superconducting and spin implementations. Some of these require weak measurements or an assumption of stationarity, but it is possible to violate the inequality with true negative-result measurements, with allowance made for possible venality due to imperfect initialization [8]. The test has been extended to higher dimensional Hilbert space, with projective measurements that are always undetectable within the protocol [9].

(iv) The Kochen–Specker theorem has been directly tested in an experiment with single photonic qutrits to show that no non-contextual theory can exist [10]. Quantum teleportation has been demonstrated over 143 km [11]. This uses entanglement and illustrates how the techniques required for quantum foundations and for quantum technologies coincide to a remarkable degree. Perhaps this is not surprising: for these very different motivations both require the degree of 'quantumness' to be experimentally extended.

## (c) Philosophy

Philosophy of quantum physics came of age with the growth in foundational studies since the 1960s. Naturally, it has focused on the 'paradoxes' of non-locality and measurement. Apart from re-evaluating, and deepening, some existing interpretations, in particular the Copenhagen and Everett interpretations, it has also assessed, from a conceptual perspective: the heterodox theories, such as the pilot-wave theory and dynamical reduction models; and the various developments in quantum information science.

(i) Over the past decade, there has been progress in constraining hidden-variable interpretations of quantum theory, i.e. interpretations that postulate physical states underlying the quantum states. Such psi-epistemic theories aim to account for randomness in measurement outcomes in terms of underlying statistical distributions of these postulated states, whereas a psi-ontic theory would allow each physical state to correspond to only one quantum state. Within Leggett–Garg concepts of macrorealism, a distinction can be made between measurements which are non-disturbing of the quantum state and measurements which are non-invasive of the physical states. The psi-epistemic models have become subject to growing constraints. It has been suggested that under certain assumptions psi-epistemic models may be ruled out or restricted [12].

(ii) There has also been progress in developing and in assessing the Everettian interpretation of quantum mechanics. In particular, Everettians have sharpened: (a) their appeal to decoherence to describe the 'splitting of worlds' as an effective process which does not conflict with relativity; (b) their appeal to decision theory to justify their applying the idea of probability, indeed the Born rule, in a multiverse in which 'everything happens' [13]. As regards assessment of the Everettian interpretation (including (a) and (b)), the state of the art can be found in a recent anthology based on another Oxford conference [14].

(iii) The idea that quantum physics involves novel logical and algebraic structures goes back to the 1930s, especially Birkhoff's and von Neumann's seminal 1936 paper 'The logic of quantum mechanics' [15]. But in recent years, new structures have been discovered and explored, often making use of category theory and its sub-field, topos theory (which mathematicians developed only after 1950). One main line of research has used category theory to provide a new graphical formalism for quantum physics, especially quantum information protocols such as teleportation. This is now sufficiently developed to yield a good comparison with the Birkhoff and von Neumann proposal [16]. Another main development has been the use of toposes to give yet a third quasi-logical formulation of quantum theory [17].

## 4. Physics today and tomorrow

We will end by setting the Oxford Questions in the context of the discussion in §1. First, we will relate them to two clouds (§4a). Finally, in §4b, we will take an even broader view, by comparing the present situation in physics with the sixteenth century scientific revolution.

### (a) Two clouds on the horizon

Our first cloud is the quantum measurement problem: that is, the difficulty of explaining completely, in terms of quantum theory, the emergence of a classical world, i.e. a world so accurately described by classical physics with its definite values—a world free of superposition and entanglement.

This cloud gets better defined by several of the Oxford Questions, as follows:

— the issue whether or not the 'collapse of the wave packet' is a physical process bears upon several Oxford Questions: in particular, 1b, 2a, 2c, 3a, 3c and 5a;
— the issue whether ideas from information theory can illuminate our concept of reality bears upon the questions: 1c, 4a, 5a and 5b; and
— the consideration of heterodox alternatives to quantum theory bears upon the questions: 1a, 2a, 3a and 4b.

Our second cloud is the search for a quantum theory of gravity. In discussions of quantum foundations, this is of course the proverbial elephant in the room (with an equally proverbial cat representing the first cloud!). It is articulated by Oxford Question 4c: how does quantum physics cohere with space–time and with mass–energy?

That is, general relativity and quantum theory are yet to be reconciled. While we have developed successful quantum theories of the other fundamental forces of Nature (electromagnetic, weak and strong), we have no analogously successful quantum theory of gravity. Accordingly, finding such a reconciliation, perhaps unification, has become an outstanding goal of theoretical physics.

There are conceptual reasons why this goal is so elusive. The contrasting conceptual structures of the 'ingredient' theories, and the ongoing controversies about interpreting them, make for conflicting basic approaches to quantum gravity. Whereas relativity theory is grounded on principles which are reasonable from a physical point of view, such as the principles of relativity and of equivalence, it remains an open question whether quantum theory could be based on comparable principles. More specifically, our first cloud, the quantum measurement problem, or the 'collapse of the wave packet', appears here in a cosmological context. How do quantum fluctuations in the early Universe, thought to be the source of gravitational perturbations that seed large-scale structures, become classical?

But we want here to emphasize another reason: namely, a dire lack of experimental data! For there are general reasons to expect data characteristic of quantum gravity to arise only in a regime of energies so high (correspondingly, distances and times so short) as to be completely inaccessible to us. To put the point in terms of length, the value of the Planck length which we expect to be characteristic of quantum gravity is around $10^{-35}$ m. This is truly minuscule: the diameters of an atom, nucleus, proton and quark are, respectively, about $10^{-10}$, $10^{-14}$, $10^{-15}$ and $10^{-18}$ m. So the Planck length is as many orders of magnitude from (the upper limit for) the diameter of a quark as that diameter is from our familiar scale of a centimetre!

### (b) Halfway through the woods

To complete this 'snapshot' of the present state of physics, we would like to endorse an analogy of Rovelli's [18]. He suggests that our present situation is like that of the mechanical philosophers, such as Galileo and Kepler of the early seventeenth century. Just as they struggled with the clues given by Copernicus and Brahe, en route to the synthesis given by Newton, so also we are 'halfway through the woods'. Of course, we should be wary of too grossly simplifying and periodizing the scientific revolution, and *a fortiori* of facile analogies between different historical situations. Nevertheless, it is striking what a 'mixed bag' the doctrines of figures such as Galileo and Kepler turn out to have been, from the perspective of the later synthesis. For all their genius, they appear to us (endowed with the anachronistic benefits of hindsight) to have been 'transitional figures'. One cannot help speculating that to some future reader of twentieth century physics, enlightened by some future synthesis of general relativity and quantum theory, the efforts of the last few decades in quantum gravity will seem strange: worthy and sensible from the authors' perspective (one hopes), but a hodge-podge of insight and error from the reader's!

Acknowledgements. We thank the participants at the conference for helping to develop and refine the Oxford Questions.

Funding statement. Sponsored by the John Templeton Foundation.